\documentclass[a4paper]{jpconf}
\usepackage{graphicx,url,lineno}
\pdfoutput=1

\newcommand{\ttbar}{\mbox{$t\bar{t}$}}
\newcommand{\ppbar}{\mbox{$p\bar{p}$}}
\newcommand{\qqbar}{\mbox{$q\bar{q}$}}

\newcommand{\etmiss}{\mbox{$E_T^{\rm miss}$}}
\newcommand{\mtt}{\mbox{$m_{t\bar{t}}$}}
\newcommand{\betatt}{\mbox{$\beta_{t\bar{t}}$}}
\newcommand{\alep}{\mbox{$A_{\rm lep}$}}
\newcommand{\ctp}{\mbox{$\cos\theta_+$}}
\newcommand{\ctm}{\mbox{$\cos\theta_-$}}
\newcommand{\dphill}{\mbox{$\Delta\phi_{\ell\ell}$}}
\newcommand{\fsm}{\mbox{$f_{\rm SM}$}}

\newcommand{\sxwt}{$\sqrt{s}=7$\,TeV}
\newcommand{\sxvt}{$\sqrt{s}=8$\,TeV}

\newcommand{\arxiv}[3]{#1, [arXiv:#3]}
\newcommand{\prd}[6]{#1, Phys.\ Rev.\ D {\bf #3} #5 (#4), [arXiv:#6]}
\newcommand{\prl}[6]{#1, Phys.\ Rev.\ Lett.\ {\bf #3} #5 (#4), [arXiv:#6]}
\newcommand{\plb}[6]{#1, Phys.\ Lett.\ B {\bf #3} #5 (#4), [arXiv:#6]}
\newcommand{\jhep}[6]{#1, J.\ High Energy Phys.\ {\bf #3} #5 (#4), [arXiv:#6]}

\newcommand{\jinst}[5]{#1, JINST {\bf #3} #5 (#4)}

\newcommand{\aconfref}[3]{ATLAS Collaboration, ATLAS-CONF-#2, 
\url{http://cdsweb.cern.ch/record/#3}}
\newcommand{\cconfref}[3]{CMS Collaboration,  CMS PAS #2, 
\url{http://cdsweb.cern.ch/record/#3}}

\newcommand{\doublefigure}[6]{
\begin{figure}[tp]
\centering
\includegraphics[width=#2]{#1}
\includegraphics[width=#4]{#3}
\caption{\label{#5}#6}
\end{figure}
}

\begin{document}
\title{Top quark property measurements at the LHC}

\author{Richard Hawkings\footnote{On behalf of the ATLAS and CMS Collaborations}}

\address{CERN, PH department, CH-1211 Geneva 23, Switzerland}

\ead{richard.hawkings@cern.ch}


\begin{abstract}
Measurements of top quark properties performed at the Large Hadron Collider
are reviewed, with a particular emphasis on top-pair charge asymmetries, 
spin correlations and polarisation measurements performed by the ATLAS 
and CMS collaborations. The measurements are generally in good agreement with
predictions from next-to-leading-order QCD calculations, and no deviations 
from Standard Model expectations have been seen.
\end{abstract}

\section{Introduction}

The top quark ($t$) is the heaviest known elementary particle, with a mass
that is much higher than those of all the other quarks, and close to the
masses of the $W$, $Z$ and Higgs bosons. Its large mass means that it decays
quickly without forming hadrons, offering the unique opportunity to study the 
properties of a `bare' quark. With its $O(1)$ Yukawa coupling to the Higgs 
boson, it may also be closely connected to electroweak symmetry breaking, or 
offer a window to physics beyond the Standard Model (SM).
Studies of the properties of the top quark are therefore central to both the 
LHC and Tevatron physics programs.

This review focuses on measurements of the top-pair (\ttbar) production
charge asymmetry, and measurements of spin correlation and top polarisation
in \ttbar\ events, performed by ATLAS \cite{atlasdet} and CMS \cite{cmsdet}
in $\sqrt{s}=7$--8\,TeV $pp$ collisions at the CERN Large Hadron Collider.
Other properties of the top quark are covered in other reviews presented
to the Top2014 conference.

\section{Top charge asymmetry}

One of the most intriguing legacy results from the Tevatron is the \ttbar\ 
forward-backward asymmetry. Analyses of the angular distributions of top
quarks and anti-quarks by CDF and D0 indicate that the top quarks (antiquarks)
are produced preferentially following the direction of the proton (antiproton)
beam. The latest measurements from CDF suggest that the size of this asymmetry 
is slightly larger than expected in the SM, whilst the measurements
from D0 are in agreement with the SM \cite{ajung}. 
This asymmetry cannot be measured at the LHC, as it collides $pp$ and 
not \ppbar, but an analogous \ttbar\ charge asymmetry $A_C$ can be defined
by looking at the difference in absolute rapidity values of the produced
top quark and antiquark. The rapidity difference 
$\Delta|y|=|y_t|-|y_{\bar{t}}|$ is positive when the top quark is produced
at a smaller angle to the beam direction (large $|y|$) than the antiquark,
and negative otherwise. The asymmetry is defined from event counts $N$ as:
\[
A_C=\frac{N(\Delta|y|>0)-N(\Delta|y|<0)}{N(\Delta|y|>0)+N(\Delta|y|<0)} .
\]
In the SM, this asymmetry is slightly positive for \ttbar\
pairs produced via $\qqbar\rightarrow\ttbar$, where interference effects
generate a correlation between the direction of the incoming quark and the
outgoing top quark, but zero for $gg\rightarrow\ttbar$. The total resulting
asymmetry is an order of magnitude smaller than the Tevatron
forward-backward asymmetry,  {\em e.g.} an NLO QCD 
calculation including electroweak corrections gives $A_C=0.0115\pm 0.0006$
at \sxwt\ \cite{ttasymqcd}.

The most precise measurements of $A_C$ come from the semileptonic \ttbar\ 
final state, in which the $W$ boson from one top quark decays to an electron
or muon and a neutrino, and the other to a \qqbar\ pair. By selecting events
with an isolated electron or muon, missing transverse momentum (\etmiss) and 
at least four jets, ATLAS and CMS both isolate samples of about 60\,000 events
with about 20\,\% non-\ttbar\ background, dominated by $W$+jets and single
top production. Kinematic fits are used to fully reconstruct the \ttbar\
system and determine $\Delta|y|$ on an event-by-event basis. The $\Delta|y|$ 
distribution is then unfolded to correct for background, efficiency
and resolution effects, and the inclusive $A_C$ corrected back to the
parton level extracted. The measurements from ATLAS \cite{atlasacljetw}
and CMS \cite{cmsacljetw} at \sxwt\ are shown in Table~\ref{t:asym}, and
have been combined \cite{ttasymcomb} to give a value of
$A_C=0.005\pm 0.007\pm 0.006$, consistent with both zero and the SM
prediction. CMS has also measured $A_C$ at \sxvt\ using a very similar
analysis \cite{cmsacljetv}.

\begin{table}
\centering

\begin{tabular}{l|l|lll|l}\hline
$\sqrt{s}$ & Asymmetry (\%) & ATLAS & CMS & LHC comb. & Theory \\
\hline
7\,TeV & semilept. $A_C$ & $0.6\pm 1.0\pm 0.5$ & $0.4\pm 1.0\pm 1.1$ & $0.5\pm 0.7\pm 0.6$ & $1.15\pm 0.06$ \\
 & dilepton $A_C$ & $2.1\pm 2.5\pm 1.7$ & $-1.0\pm 1.7\pm 0.8$ & & $1.15\pm 0.06$ \\
 & dilepton \alep & $2.4\pm 1.5\pm 0.9$ & $0.9\pm 1.0\pm 0.6$ & & $0.70\pm 0.03$ \\ \hline
8\,TeV & semilept. $A_C$ & & $0.5\pm 0.7\pm 0.6$ & & $1.11\pm 0.04$ \\
\hline
\end{tabular}
\caption{\label{t:asym} Measurements of the inclusive \ttbar\ asymmetry $A_C$
and leptonic asymmetry \alep\ from ATLAS \protect\cite{atlasacljetw,atlasacllw}
and CMS \protect\cite{cmsacljetw,cmsacljetv,cmsacllw} at \sxwt\ and \sxvt\ with
their statistical and systematic uncertainties, together with the LHC 
combination \protect\cite{ttasymcomb} and corresponding theoretical predictions
\protect\cite{ttasymqcd}.}
\end{table}

Both collaborations have also measured $A_C$ differentially, as a function
of the rapidity, transverse momentum and invariant mass of the \ttbar\ system.
The latter in particular increases the sensitivity to new physics
scenarios, which are expected to be more prominent at high \mtt\ as shown
in Figure~\ref{f:acljets}. ATLAS has also measured the asymmetry for 
$\betatt>0.6$, this cut on the longitudinal velocity of the \ttbar\ system 
increasing the fraction of $\qqbar\rightarrow\ttbar$ events and the potential
new physics contributions, but no significant deviations from the SM have been 
seen.

\doublefigure{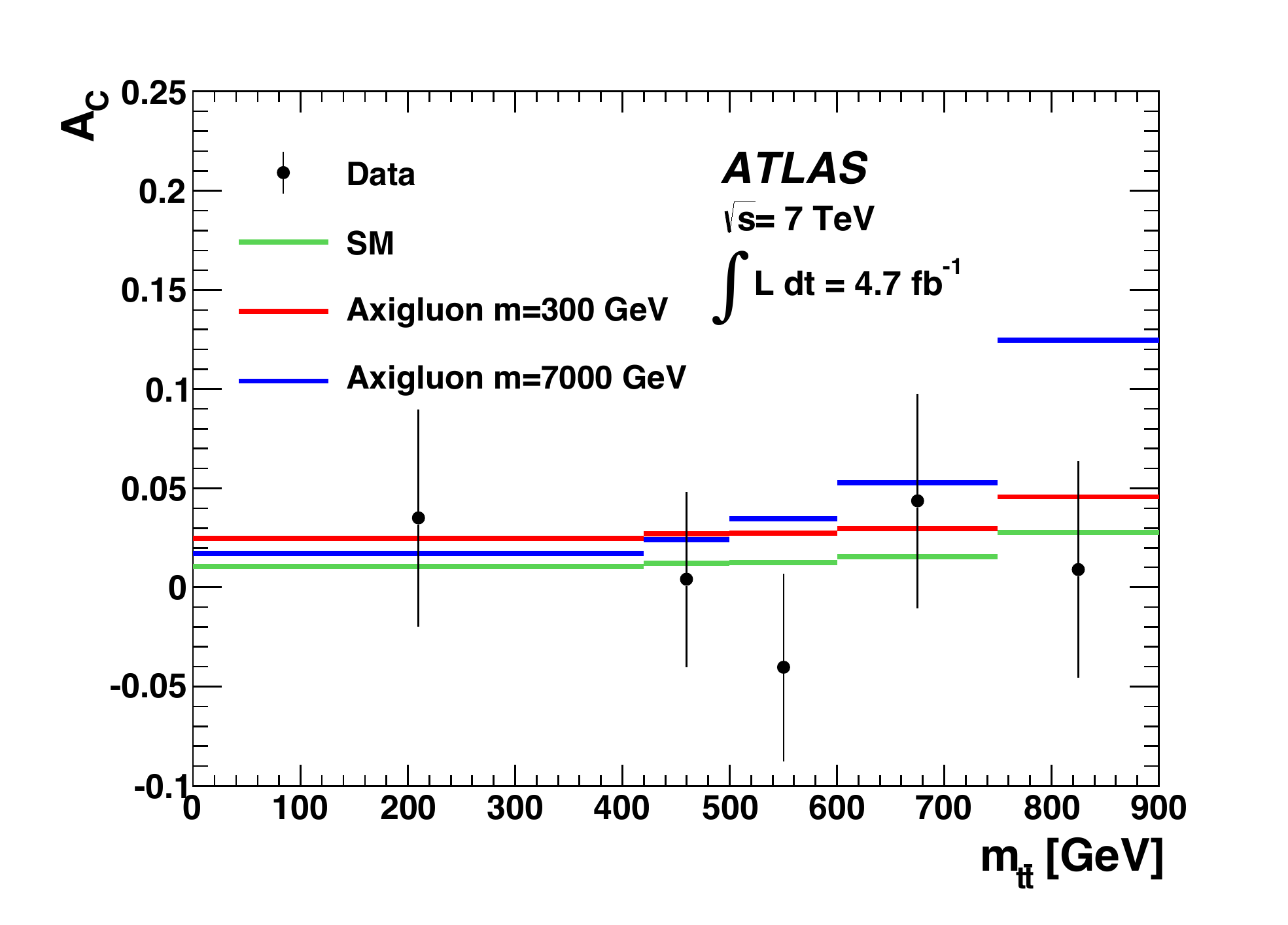}{70mm}{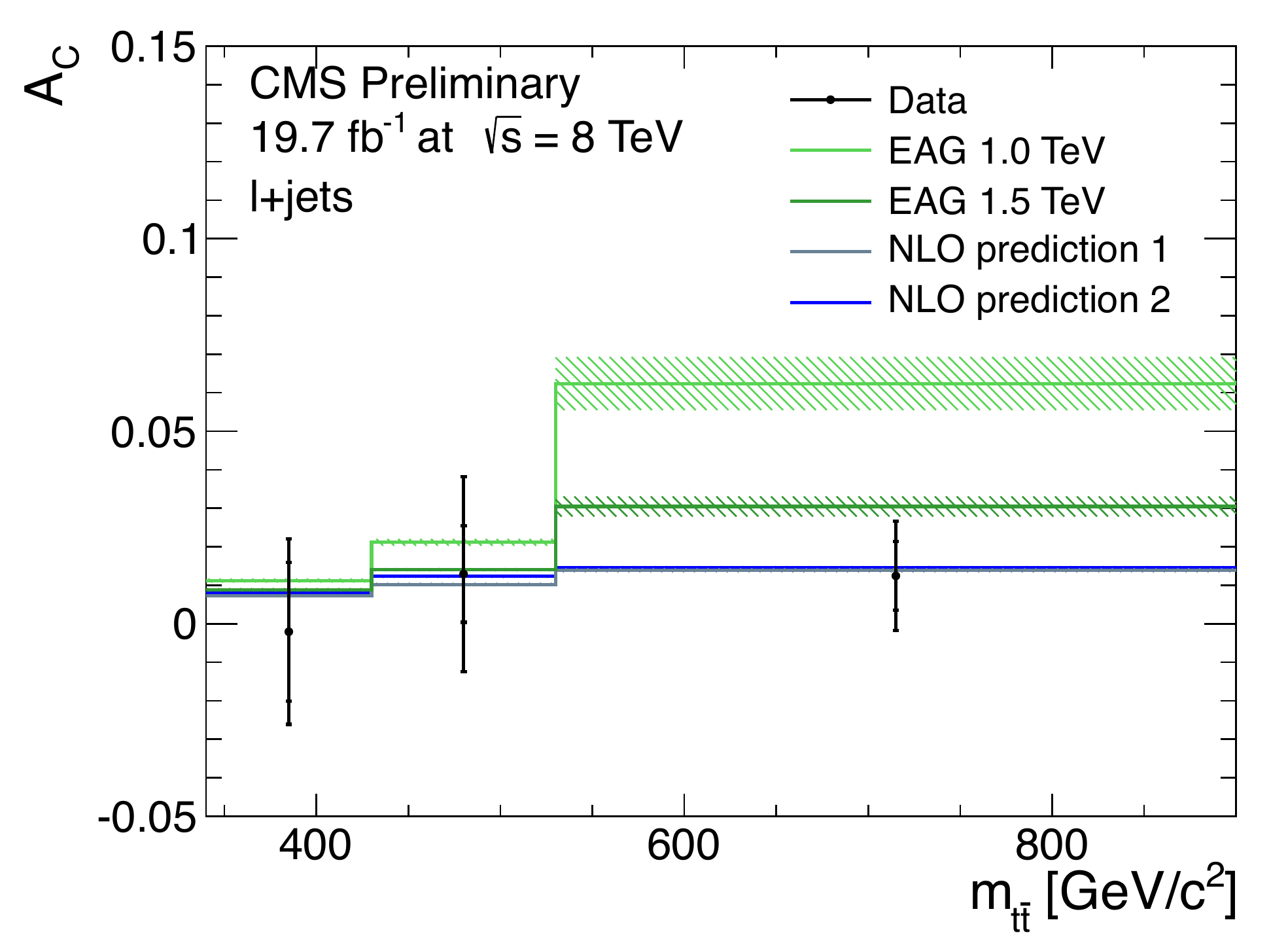}{65mm}{f:acljets}
{Measurements of the \ttbar\ charge asymmetry $A_C$ as a function of the
invariant mass of the \ttbar\ system in semileptonic \ttbar\ events
from ATLAS at \sxwt\ \protect\cite{atlasacljetw} and CMS at \sxvt\ 
\protect\cite{cmsacljetv}.
The measurements are compared to SM NLO QCD predictions
with electroweak corrections from Ref. \cite{ttasymqcd} (CMS NLO1) and
Ref. \protect\cite{ttasymsi} (CMS NLO2 and ATLAS), and to various 
beyond-Standard Model scenarios (see \protect\cite{atlasacljetw,cmsacljetv} 
for details).}

The \ttbar\ charge asymmetry has also been measured in the dilepton
channel at \sxwt\ \cite{cmsacllw,atlasacllw}. In 
dileptonic \ttbar\ events, the $W$ bosons from both top quarks decay to
leptons and neutrinos, giving an under-constrained system since the
\etmiss\ measurement cannot resolve the separate contributions of each 
neutrino. Extra assumptions, such as the expected distribution of neutrino
rapidities, are used to find the most probable kinematic configuration
for each event, allowing $\Delta|y|$ to be extracted and $A_C$ to 
be measured. In dileptonic events, a complementary asymmetry observable \alep\ 
can be defined, based on the difference in $|\eta|$ between the positive
and negatively charged leptons: $\Delta|\eta|=|\eta^{\ell +}|-|\eta^{\ell -}|$.
The distributions of both variables for the updated ATLAS dilepton
analysis \cite{atlasacllw} (new for this conference) are shown in 
Figure~\ref{f:aclldist}, and the $A_C$ and \alep\ values from both
ATLAS and CMS are shown in Table~\ref{t:asym}.

\doublefigure{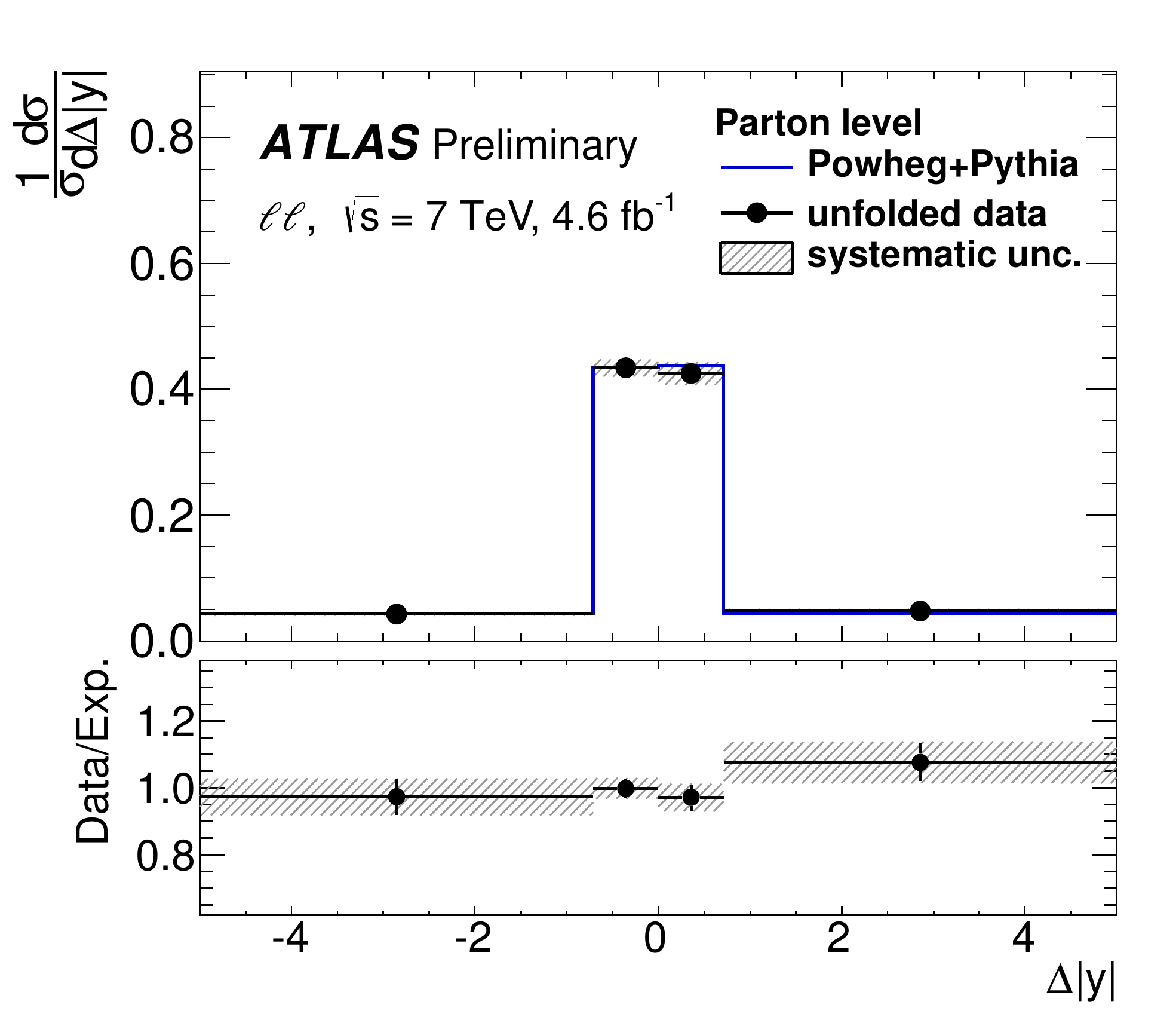}{61mm}{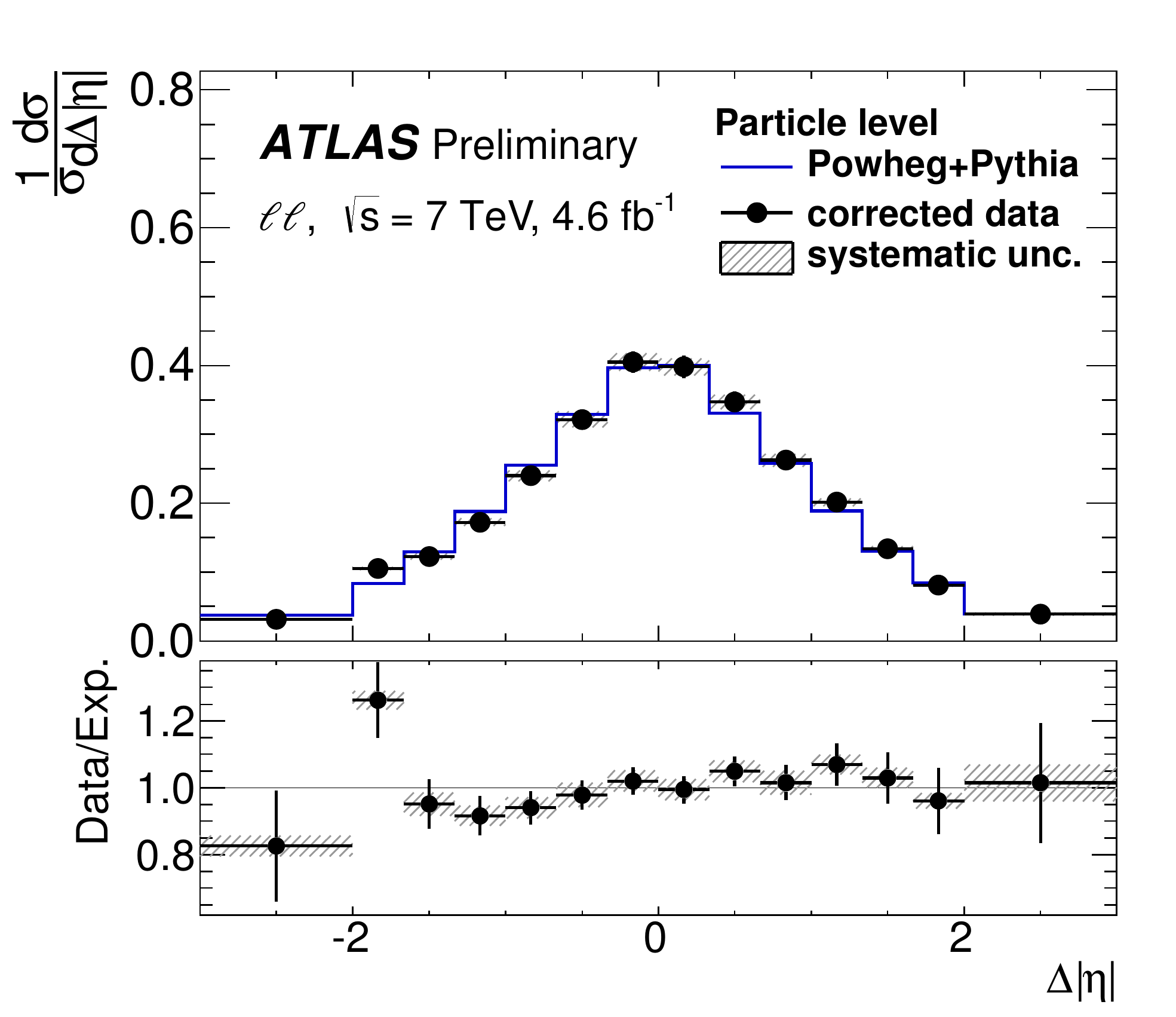}{61mm}{f:aclldist}
{Unfolded normalised \ttbar\ $|\Delta|y|$ distribution (left) and dilepton
$\Delta|\eta|$ distribution (right) from the ATLAS dilepton
charge asymmetry analysis at \sxwt\ \protect\cite{atlasacllw}. The data is 
compared to the simulation predictions based on {\sc Powheg+Pythia}.}

The inclusive charge asymmetry measurements are summarised in 
Table~\ref{t:asym}. The semileptonic measurements have a precision of around 
1\,\%, not yet precise enough to distinguish between zero and the small
non-zero asymmetry expected in the SM. The dilepton measurements have
lower precision, due to the smaller event samples and ambiguities inherent
in the dileptonic final state. 


\section{Spin correlations and polarisation}

The top quark lifetime of about $3\times 10^{-25}$\,s is much shorter than the
time required to form hadrons, so the top quark decays as a `bare' quark,
transferring information on its spin to the decay products. In \ttbar\
pair production, the polarisation is expected to be negligible, but the
spins of the $t$ and $\bar{t}$ are correlated, such that the asymmetry
in the numbers of events where the $t$ and $\bar{t}$ have like and unlike
spins, $A=(N_{\rm like}-N_{\rm unlike})/(N_{\rm like}+N_{\rm unlike})$, is non-zero.
The double-differential cross-section in the decay angles of the two 
top quarks, \ctp\ and \ctm, is given by:
\[
\frac{1}{\sigma}
\frac{{\rm d}\sigma}{{\rm d}\ctp {\rm d}\ctm}=\frac{1}{4}(1+A\alpha_+\alpha_-
\ctp\ctm)
\]
where the asymmetry parameter $A$ is scaled by the spin-analysing power
of the chosen top quark decay products, and \ctp\ and \ctm\
are the cosines of the angles between the top quark decay product and the chosen
polarisation axis. Normally the helicity basis is used, in which the 
polarisation is measured using the direction of the top quark momentum in the
\ttbar\ rest frame. The spin-analysing power $\alpha$ is $0.998$ for
positively-charged leptons, $-0.966$ for down quarks from the $W$ decay, and
-0.393 for $b$ quarks from the top decay.

Measurement of the \ttbar\ spin correlations using the above formalism 
requires the full reconstruction of the \ttbar\ system, using techniques
similar to those used in the charge asymmetry measurement. They have been
measured at \sxwt\  by CMS in dilepton events \cite{cmsspinpol} 
(see Figure~\ref{f:cmspol} (left)), and by  ATLAS in both dileptonic and 
semileptonic events \cite{atlasspdphiw}, using both the helicity basis and
the so-called maximal basis which is particularly sensitive to 
correlations in the $gg\rightarrow\ttbar$ subprocess. All measurements are
consistent with the spin correlations predicted by the Standard Model.

\doublefigure{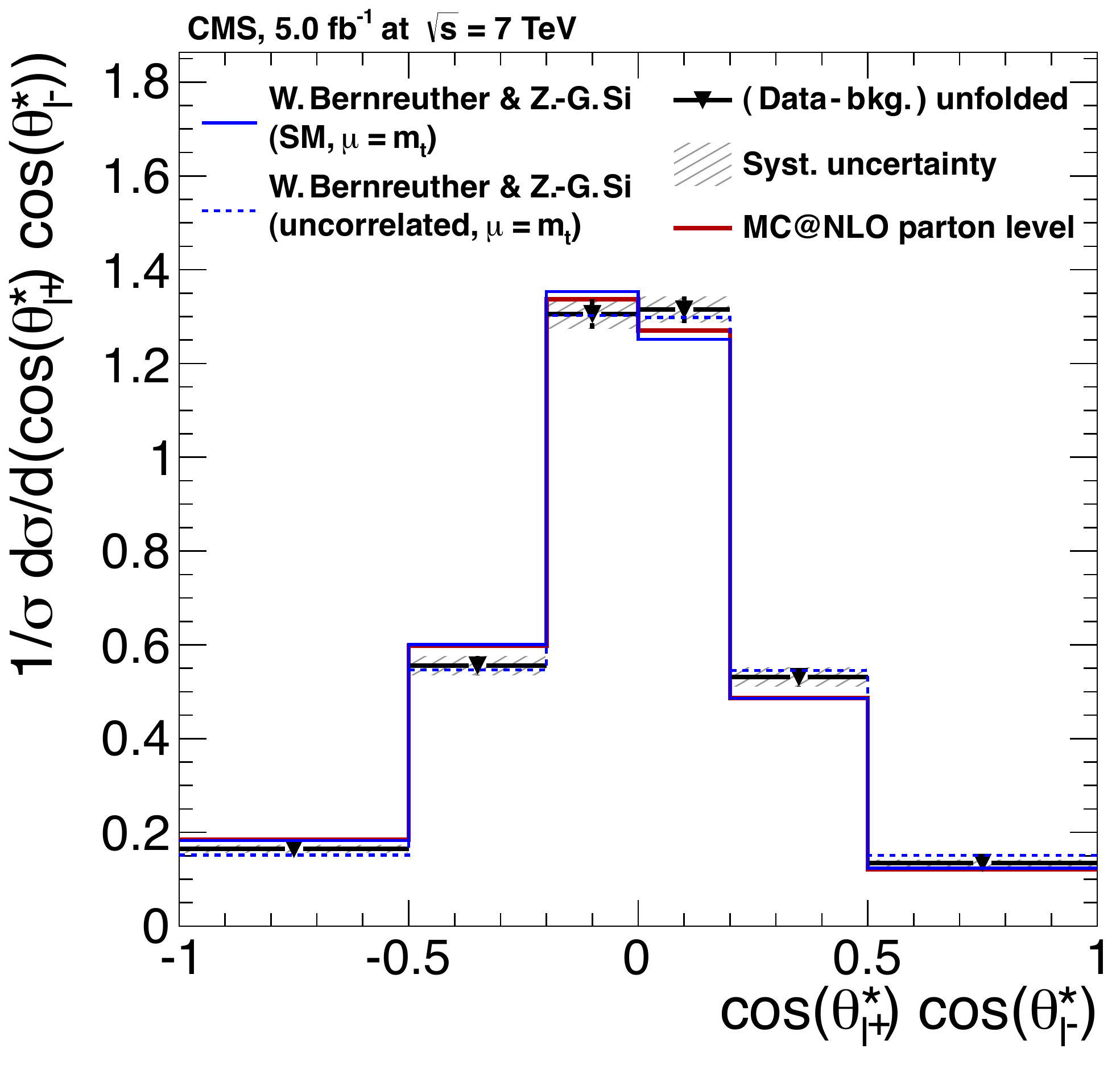}{55mm}{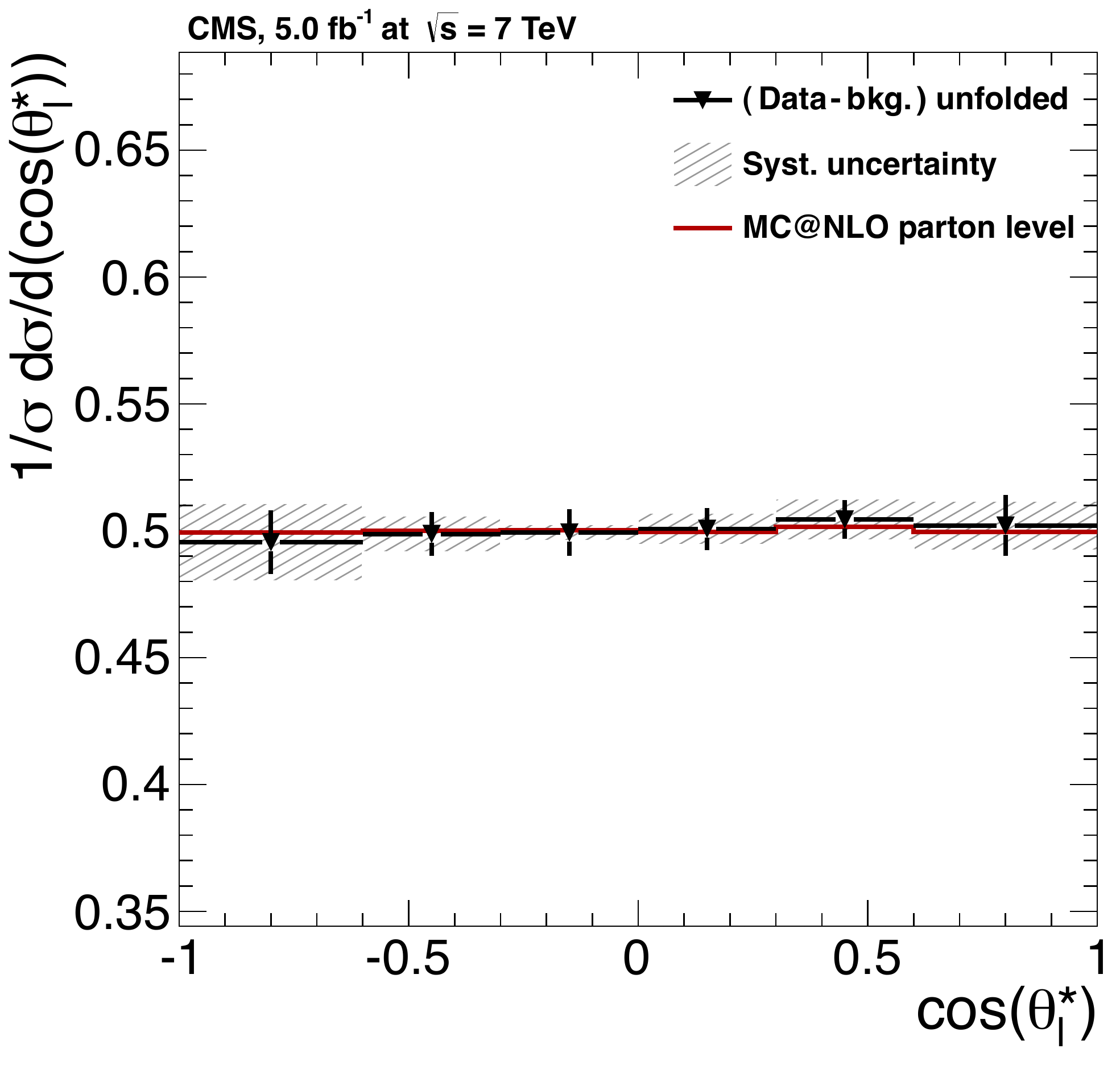}{55mm}{f:cmspol}{Unfolded
distributions of $\ctp\ctm$ (left) and top quark decay angle $\cos\theta$ 
(right) measured by CMS in dilepton \ttbar\ events at \sxwt\ 
\protect\cite{cmsspinpol},
and compared to predictions with and without \ttbar\ spin correlations
and from the MC@NLO \ttbar\ event generator.}

The \ttbar\ spin correlation also affects the distribution of \dphill, the
difference in azimuthal angles (transverse to the beamline) of the two leptons
in dileptonic \ttbar\ events. Both collaborations exploited this variable
to demonstrate the existence of spin correlations in \sxwt\ data. The CMS
analysis \cite{cmsspinpol} unfolded the \dphill\ distribution to parton
level, correcting for background, acceptance and resolution effects, and
compared the result to simulation predictions with and without spin correlations
as predicted by the SM (see Figure~\ref{f:spdphi} (left)). 
The level of spin correlation was quantified from the asymmetry of the
parton-level \dphill\ distribution about $\dphill=\pi/2$ as 
$A_{\Delta\phi}=0.113\pm 0.010\pm 0.013$, in agreement with the NLO
prediction of 0.115.
ATLAS \cite{atlasspdphiw} instead compared the detector-level \dphill\ 
distribution to fully-simulated events with and without spin correlation,
quantifying the spin correlation strength with a fit to SM-like correlated and
uncorrelated simulation-derived templates. The fitted fraction of correlated 
template \fsm\ was measured to be $1.19\pm 0.09\pm 0.18$, compatible with unity.
At the conference, ATLAS presented a new analysis using the \dphill\ 
distribution at \sxvt\ \cite{atlasspdphiv} with an optimised dilepton
event selection having higher efficiency; as shown in 
Figure~\ref{f:spdphi} (right), this is also compatible with the SM 
expectation, with a fitted $\fsm=1.20\pm 0.05\pm 0.15$, corresponding
to a spin correlation strength in the helicity basis of $A=0.38\pm 0.04$.
The \dphill\ distribution can also be used to set limits on new physics
contributions within the selected \ttbar\ event sample,  for example
top squark pair production as shown by ATLAS \cite{eifert}, 
or an  anomalous \ttbar-gluon interaction parameterised as a chromomagnetic
dipole moment as explored by CMS \cite{cmschromo}.

\doublefigure{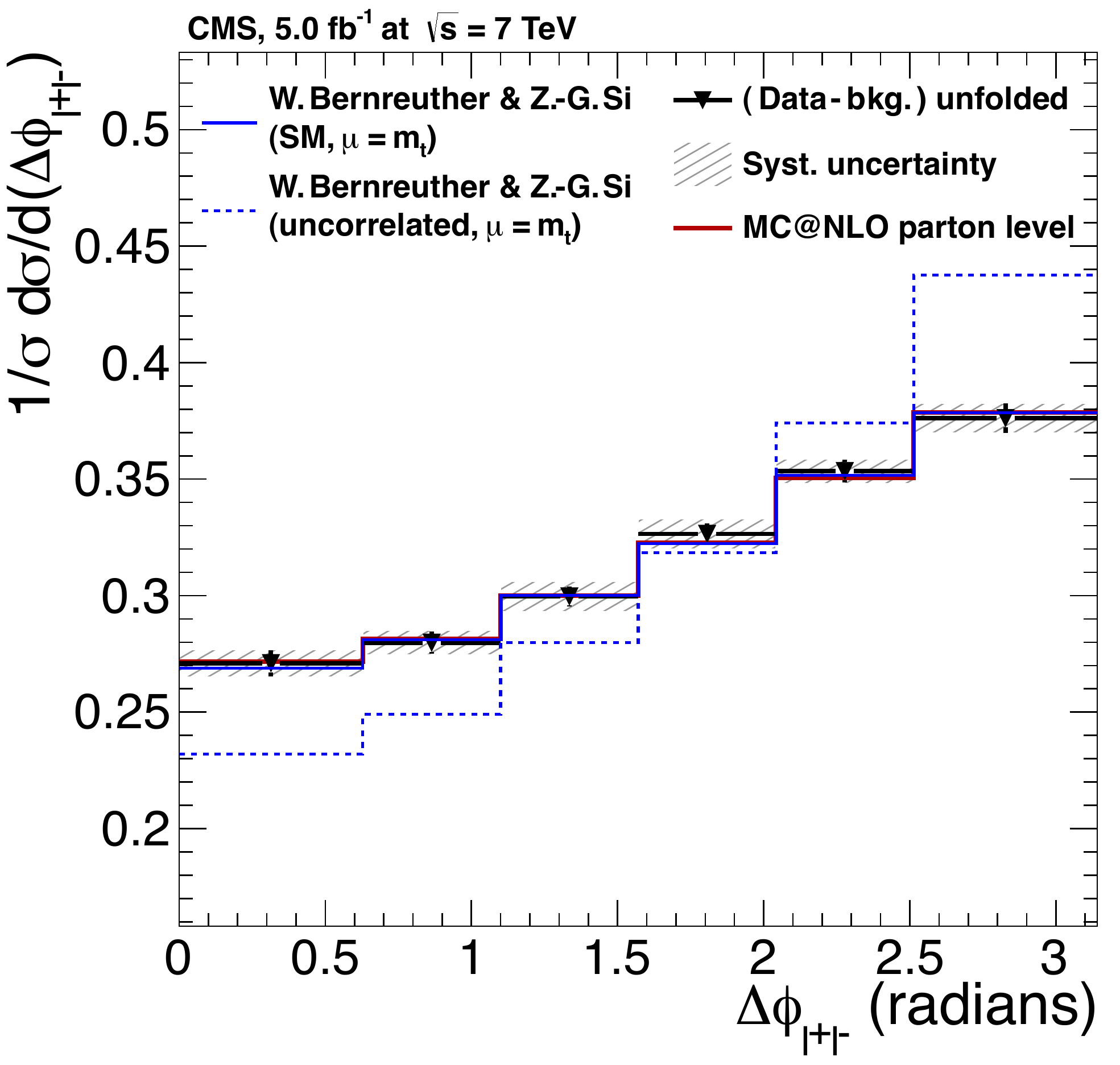}{60mm}{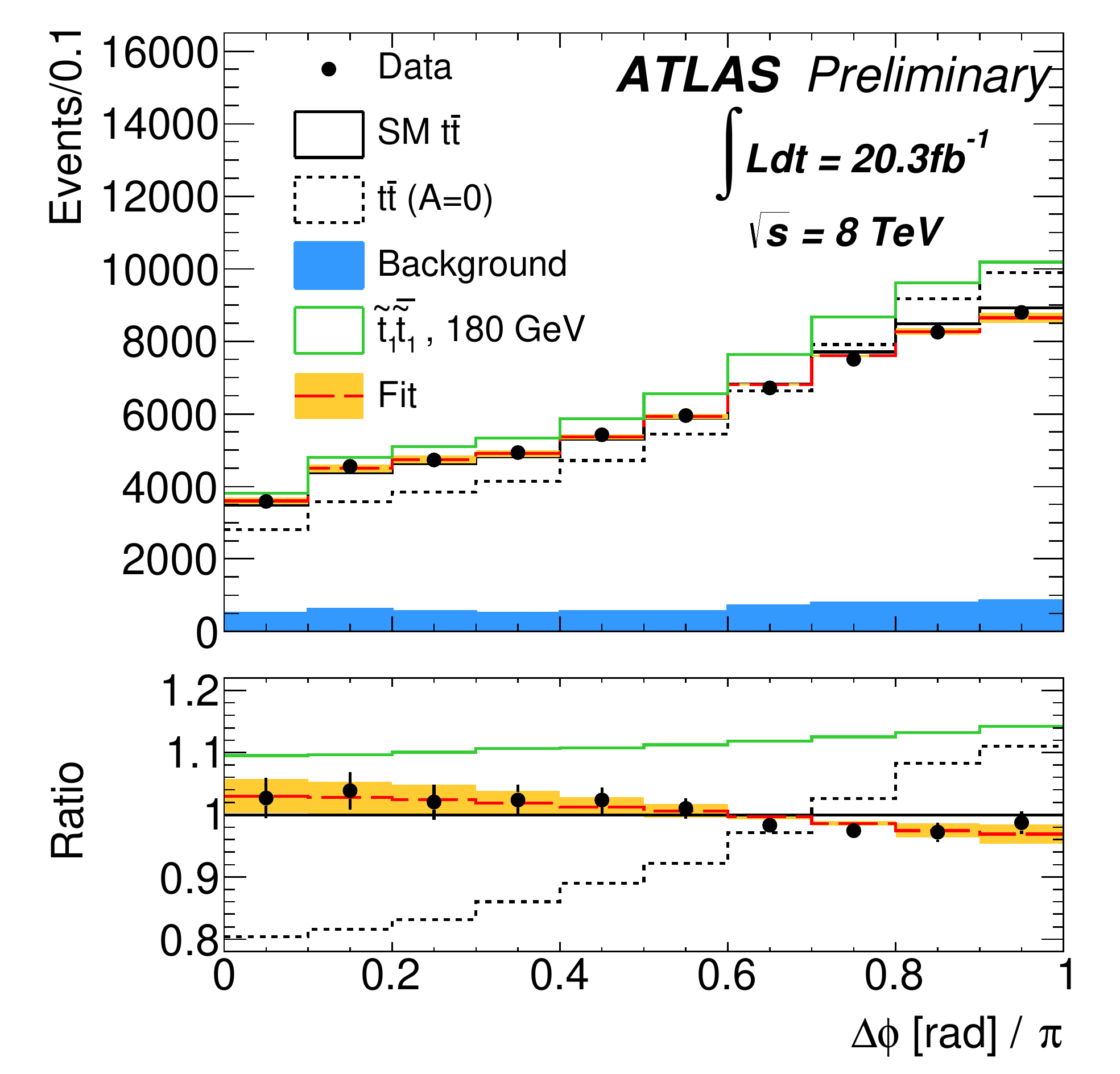}{60mm}{f:spdphi}
{Measurements of spin correlation using the \dphill\ distribution in 
dilepton \ttbar\ events from CMS at \sxwt\ \protect\cite{cmsspinpol} 
and ATLAS at \sxvt\ \protect\cite{atlasspdphiv}, compared to simulation
predictions with and without spin correlation at parton level (CMS) or 
detector level (ATLAS).}

The CMS collaboration also used the same \sxwt\ dilepton sample to measure
the top quark polarisation in \ttbar\ events, via the unfolded distribution
of the decay angle $\cos\theta$ \cite{cmsspinpol}. The results are shown in 
Figure~\ref{f:cmspol} (right), and are consistent with zero polarisation. ATLAS
also measured the top polarisation in both dilepton and semileptonic events,
looking for polarisation effects which polarise the $t$ and $\bar{t}$ quarks
with the same (CP-conserving) or opposite (CP-violating) signs, and again
found results compatible with zero \cite{atlaspol}.

\section{Outlook}

The LHC data have already produced a wealth of top quark property measurements, 
so far mainly at \sxwt. Top-pair charge asymmetry measurements are reaching
1\,\% precision, and the inclusion of the full \sxvt\ dataset may allow
the small non-zero expected asymmetry to be observed. No hints of deviations
in either the inclusive or differential asymmetries have been seen, strongly
constraining various new physics scenarios. The expected \ttbar\ spin 
correlations have been clearly observed at both \sxwt\ and \sxvt, and 
the top quark polarisation in \ttbar\ events has been seen to be consistent
with the expectation of zero within about 2\,\%. Many of these measurements
are limited by systematics, and advances in \ttbar\ modelling or new 
analysis techniques will be needed to fully exploit the \sxvt\ dataset. This
will be even more important in the upcoming $\sqrt{s}=13$--14\,TeV run, which
in particular will allow top charge asymmetry measurements to be
pushed to sub-percent precision and extended to higher \ttbar\ invariant masses.


\end{document}